# Passive, Noiseless, Intensity Amplification of Repetitive Signals

Reza Maram[1], James Van Howe[1,2], Ming Li[1,3] & José Azaña[1,*]

[1]*Institut National de la Recherche Scientifique (INRS) – Energie, Matériaux et Télécommunications, Montréal, Québec, Canada, H5A 1K6, Tel.: +1 (514) 875 1266 – 7018  Fax: +1 (514) 875 0344.*

[2]*Department of Physics and Astronomy, Augustana College, Rock Island 61201, USA*

[3]*State Key Laboratory on Integrated Optoelectronics, Institute of Semiconductors, Chinese Academy of Sciences, Beijing 100083, China.*
*Correspondence to: azana@emt.inrs.ca

**Amplification of signal intensity is essential for initiating physical processes, diagnostics, sensing, communications, and scientific measurement. During traditional amplification, the signal is amplified by multiplying the signal carriers through an active gain process using an external power source. However, for repetitive waveforms, sufficient energy for amplification often resides in the signal itself. In such cases, the unneeded external power is wasted, and the signal is additionally degraded by noise and distortions that accompany active gain processes. We show noiseless intensity amplification of repetitive optical pulse waveforms with gain from 2 to ~20** *without using active gain*, **by recycling energy already stored in the input repetitive signal. This "green" method uses dispersion-induced temporal self-imaging (Talbot) effects to precisely re-distribute the original signal energy into fewer replica waveforms. This approach simply requires a suitable manipulation of the input signal's phase profile along the temporal and spectral domains. In addition, we show experimentally how our passive amplifier performs a real-time average of the wave-train to reduce noise fluctuation present on the input pulse train, as well as enhances the extinction ratio of pulses to stand above the noise floor by approximately the passive amplification gain factor. Finally, our technique is applicable to repetitive waveforms in any spectral region or wave system, including acoustic, mechanical, and quantum probability waveforms, for which active amplification methods are challenging to implement or nonexistent.**

Waveform amplification refers to a process by which the amplitude of the waveform is increased without affecting its other signal features, particularly temporal shape. Intensity amplification of waveforms is necessary for increasing the peak power of signals for initiating physical processes, extracting information from the natural world, and communications. Almost every information-bearing electronic signal used for diagnostics[1], sensing[2], or fundamental measurement[3] requires amplification between the source and the detector. Optical pulses amplified to high-intensities are necessary for nonlinear microscopy[4], materials processing[5], and relativistic optical processes such as coherent x-ray generation, laser fusion, and particle acceleration[6]. Additionally, optical amplification of signal intensity is critical for communications[7] and becoming increasingly important for current work in optical computing[8] and optical information processing[9]. During active amplification, the signal is amplified directly by multiplying the signal carriers through an active gain process using an external power source[10-17]. Whereas active gain mechanisms are widely available for electrical and optical signals, suitable active gain processes for direct waveform amplification are limited over large regions of the electromagnetic spectrum and for other wave systems, such as mechanical, acoustic, and quantum probability waves. Typical methods of amplification for these waveforms entail the use of transducers to convert the signal into one that can be amplified directly, such as acoustical-to-electrical conversion, or in the case of quantum waveforms, through indirect methods such as entanglement[18].

Furthermore, for applications involving high-peak power pulse generation[4-6], traditional active amplification is extremely inefficient. In order to generate high-peak power optical pulses, many pulses in a wave-train are intentionally thrown away, "pulse-picking," either prior to amplification[19] or during the amplification process (regenerative amplification)[20], in order to concentrate energy from the external source in fewer remaining output pulses. In such processes, typically more than 99% of the signal energy is thrown away prior to boosting.

Finally, in all applications where active amplification is employed the active-gain process also contributes amplitude and phase noise, e.g., amplified spontaneous emission (ASE) noise or timing jitter, and other signal distortions induced by limited gain frequency bandwidth. Because an active-gain process inherently amplifies input noise and also injects its own noise onto a signal, the output signal-to-noise ratio (SNR) always degrades. This degradation is encapsulated by a common figure-of-merit of any active amplifier- the noise figure[21]. The noise figure of an active amplifier is the ratio of the SNR of the input signal to the SNR of the output signal. In the cases of weak signals, even active amplifiers with low noise-figures will render signals totally undetectable by the amplification and injection of noise.

Using optical pulses, we demonstrate a noiseless waveform amplification method for repetitive signals, with experimental intensity gain from 2 to ~20, *without using an active gain process* by recycling energy already stored in the original waveform signal. Our new concept of passive amplification exploits the intrinsic coherence revival times of periodic waveform trains provided by passive dispersive broadening, known as temporal "self-imaging" or the temporal Talbot effect[22] (illustrated in Fig. 1, top). In particular, passive amplification is achieved by re-distributing the overall input energy of a repetitive waveform signal into fewer waveforms, resulting in each output individual waveform to be an amplified copy of the input. This can be interpreted as coherently adding a desired number of identical copies on top of one another to produce the desired amplified output. For example, in one of our demonstrated experiments, we take a train of identical pulses and repeatedly add every adjacent group of 14 pulses onto the 15th pulse to create an output pulse train with 1/15th the input repetition rate and 15 times the intensity (here the passive gain factor is $m=15$). Thus, we implement a "lossless" pulse picking process by using a suitable combination of *phase-only* temporal and spectral modulation operations, in order to preserve the overall input signal energy.

Controlled coherent addition of many identical waveforms is an extremely challenging process. For instance, coherent addition of optical pulses in a wave-train has been shown in the past, but only by precisely timing and storing pulses in a high-finesse cavity[23, 24]. This method requires ultra-precise active phase control of the input signal envelope and carrier, i.e., carrier-envelope phase stabilization, in order to stabilize the input waveforms to cavity such that each pulse will add constructively. Otherwise, small phase misalignment in each pulse tends towards destructive interference. Though impressive results have been reported, these conditions are much too restrictive to be applied beyond a controlled lab environment. Using self-imaging effects, we similarly show pulse amplification by coherent addition, however, without using a cavity and therefore without the associated stringent timing and stabilization conditions. The only condition on the input waveform is that it is repetitive and maintains coherence over the output repetition rate. Because our passive amplification technique for repetitive signals is based upon the fundamental process of linear superposition of waves, it can be applied to any wave system, introducing a method for multiplying signal intensity in systems where active gain processes for waveform amplification are limited or do not exist. For example, direct passive amplification of acoustic and vibrational waveforms without the need for transducers could be useful in MEMS applications, sonar, and ultrasonic sensing and imaging. Additionally, passive amplification of probability waves could provide another tool for atom optics to control and manipulate quantum states of matter[25] as well as single photon pulses[26].

Last but not least, we report numerical and experimental results to demonstrate that during passive intensity amplification, the SNR does not degrade. In fact, the proposed Talbot-based amplification technique *improves* random (white) intensity noise, ASE-like fluctuations, present on the input temporal signal, similarly to a real-time averaging process, as well as enhances the extinction ratio compared to the input waveform train in the time-domain. In particular, consistent with simulation, we show experimentally that Talbot-based passive amplification increases the extinction ratio by approximately the passive gain factor, *m*, as well as performs a real-time average of the waveform train by about *m* averages. These noise-mitigating effects allow us to show in a side-by-side comparison in one of our experiments how Talbot-amplification can effectively extract a weak signal buried beneath a noisy background where active amplification buries the signal further beneath the noise. Initial data and simulation (not shown in this report) also suggest that Talbot amplification improves pulse-to-pulse fluctuation and timing-jitter present on the input signal as the passive gain factor increases. This is consistent with previous experiments which show improvement in timing-jitter and pulse-to-pulse fluctuation for simple self-imaging (passive gain factor of $m=1$)[27]. Noise mitigation using Talbot passive amplification therefore shows promise for further signal regeneration than the results presented in this report.

Figure 1 illustrates the concept of our passive amplification technique. We effectively exploit an "inverse temporal Talbot effect" in which the final amplified temporal image is recovered from a previous multiplied self-image. As shown in Fig. 1, top, in the standard temporal Talbot effect, a flat-phase repetitive input waveform (signal at $z=0$) is self-imaged, after dispersive propagation through a distance $z_T$ (integer Talbot distances). There also exists an infinite amount of fractional distances, given by the "Talbot Carpet" (a map of all possible coherence revivals resembling a Persian rug in intricate but repeating patterns)[22] that give multiplied self-images; see examples at the fractional Talbot distances $z_T/2$ and $2z_T/3$, Fig. 1, top. Dispersive propagation speeds up and slows down the different frequency-components 'colors', originally in-phase that make up the waveform train, redistributing the original signal energy into the mentioned different temporal intensity patterns. An integer self-image exhibits the same repetition rate and individual waveform intensity as the input, whereas in the multiplied self-images, the repetition rate is increased, and the individual waveform intensity is correspondingly decreased by an integer factor. The repetition rate-multiplication (intensity-division) factors for the multiplied self-images shown in Fig. 1 at $z_T/2$ and $2z_T/3$ are 2 and 3, respectively. In an integer self-image, the uniform temporal phase profile of the input is restored. However, in the multiplied self-images, such as those observed at distances $z_T/2$ and $2z_T/3$, there exists a waveform-to-waveform residual temporal phase structure (dashed black). This residual temporal phase represents instances where the waveform field-amplitude has been advanced or delayed in relation to the envelope center.

By using a multiplied image at a fractional distance as the input instead of the conventional phase-free input at $z=0$, further dispersive propagation to the distance $z_T$ produces an output with an amplified intensity. As shown by Fig. 1 top, a waveform starting at the fractional distance $z_T/2$ will be amplified in intensity by a factor of *m*=2 at the output $z_T$. Likewise, a waveform starting at the fractional distance $2z_T/3$ will be amplified by a factor of *m*=3. This requires the application of a prescribed temporal phase modulation profile to an input signal to make it appear as though it has already propagated from $z=0$ through an amount of dispersive delay equivalent to the target multiple image. For example, Fig. 1, bottom, shows how if we condition a typical flat-phase input to look like the waveform train at $2z_T/3$ by proper temporal phase modulation (dashed black line), subsequent propagation through $z_T/3$ more of dispersive delay will give the output shown at $z_T$, one-third the repetition rate and three times the intensity. Notice that the described processes involve only a suitable manipulation of the input signal temporal and spectral phase profiles, not magnitude, ensuring that the signal energy is ideally preserved. The full Talbot carpet provides an infinite amount of fractional self-image locations and corresponding phase profiles, so that *any* desired repetition rate division and corresponding amplification factor can be obtained, limited only practically by the control of temporal phase modulation and spectral phase filtering from dispersive propagation.

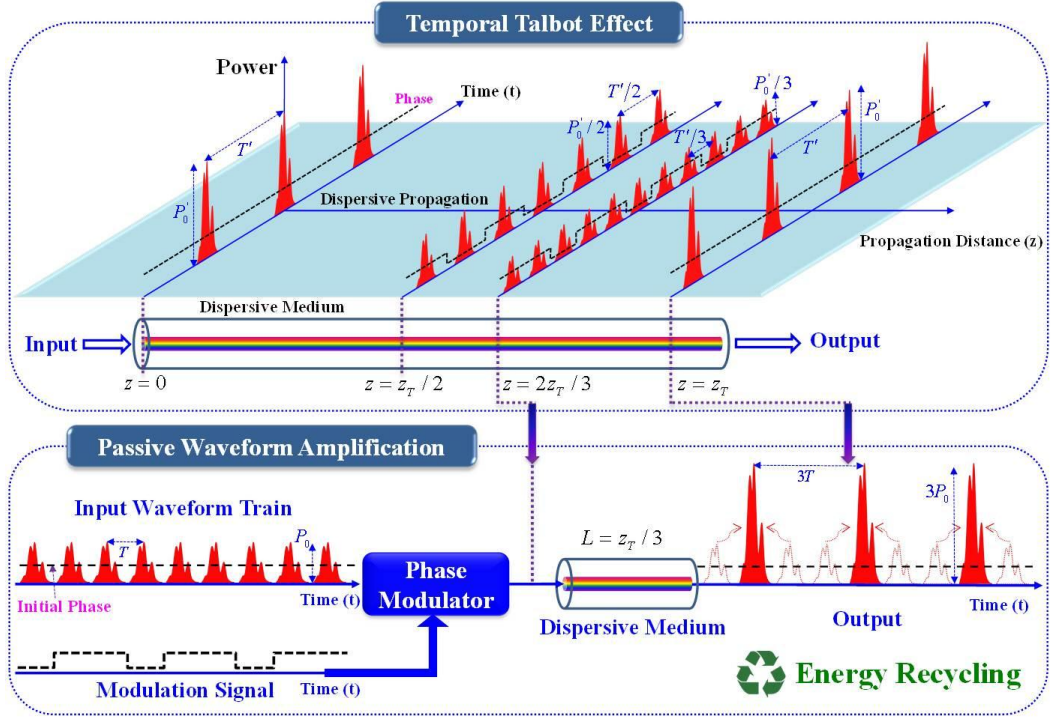

**Figure 1 | Passive waveform amplification concept.** Coherent addition of repetitive pulse waveforms is achieved by tailoring the temporal phase of the input waveform train and subsequent propagation through a dispersive medium such that the individual waveform intensity (energy) is amplified by an amount the repetition rate is reduced. A portion of the temporal Talbot carpet, top, provides the map of the temporal phase modulation and spectral phase-only filtering from dispersion required for passive amplification. In this representation, $z_T$ is the Talbot distance at which an exact (integer) "self-image" of a flat-phase incoming pulse train (signal at $z=0$), with original repetition period $T'$, is obtained by simple propagation through the dispersive medium. In the amplification example illustrated here, bottom, by applying the temporal phase variation corresponding to the multiplied Talbot self-image at distance $2z_T/3$ on an input waveform train with repetition period $T\,(=T'/3)$, subsequent propagation through $z_T/3$ of dispersive delay coherently shifts energy such that the output repetition rate is reduced by a factor of three and the waveform intensity is correspondingly enhanced by a factor of three.

To passively amplify any arbitrary repetitive input signal by $m$ times ($m=2,3,4,\ldots$), a temporal phase of

$$\phi_n = \frac{s}{m}\pi n^2 \tag{1}$$

(where $s = m-1$) is applied on the $n$-th temporal pulse ($n = 0,1,2,\ldots$) of the input periodic signal. This is followed by propagation through a first-order dispersive medium, ideally providing a linear group-delay variation as a function of frequency over the signal spectral bandwidth. The dispersive medium should introduce a total dispersion value:

$$2\pi |\beta_2| z_A = mT^2 \qquad (2)$$

where $T$ is the repetition period of the input pulse train, $z_A$ is the length of the dispersive medium, and $\beta_2$ is the dispersion coefficient, defined as the slope of the group delay as a function of the radial frequency $\omega$, per unit length. Mathematically, $\beta_2 = \left[\partial^2 \beta(\omega)/\partial \omega^2\right]_{\omega=\omega_0}$, where $\beta(\omega)$ is the propagation constant through the medium, i.e. $\tau_g(\omega) = z_A \left[\partial \beta(\omega)/\partial \omega\right]_{\omega=\omega_0}$ is the medium's group delay, and $\omega_0$ is the central (carrier) frequency of the considered signal (periodic waveform train). The temporal phase $\phi_n$ can be assumed to be applied on time slots equal to the pulse repetition rate. In practice, however, it is sufficient to apply the same phase over the pulse duration. Notice that the phase profile $\phi_n$ is periodic with a fundamental period equal to the gain-factor $m$, namely $\phi_n = \phi_{n+m}$. If these phase shifts are reduced to a $2\pi$ range, a periodic sequence of discrete phase steps in the range $[0, 2\pi]$ is obtained. The example illustrated in Fig. 1 of the main text for $m = 3$ gives a repeating temporal phase profile of $\{0, 2\pi/3, 2\pi/3, 0, 2\pi/3, 2\pi/3, \ldots\}$, where the third phase level corresponding to $n = 2$ has been obtained by taking modulo $2\pi$.

The temporal phase shifts $\phi_n$ are defined from the Talbot carpet[22]. These phase shifts induce a spectral self-imaging (Talbot) effect on the modulated pulse train[28]. In particular, the temporal phase modulation process produces new frequency components, reducing the frequency spacing of the input signal discrete comb-like spectrum by an integer factor of $m$. This is consistent with the repetition-rate division (temporal period increase) by a factor $m$ that is subsequently achieved on the temporal pulse train following dispersive propagation.

**Results**

We experimentally demonstrate passive amplification using optical pulses generated from a standard commercial pulsed fibre laser, which does not incorporate any carrier-envelope phase stabilization mechanism. The pulses from the laser are input directly into a dispersive optical fibre and delivered at the fibre output through conventional integer temporal self-imaging[22]. Next, by adding a suitable temporal phase through electro-optic modulation to waveforms prior to dispersion, the waveform intensity will be locally amplified according to the amount of repetition rate reduction. Because the absence of temporal phase modulation produces an integer Talbot self-image of the original pulse train at the fibre output, we are able to compare the cases of passively amplified and unamplified waveforms that have propagated through the same optical system. Any loss from the system will show up in both the passively amplified and unamplified data, allowing us to isolate the effectiveness of passive amplification alone.

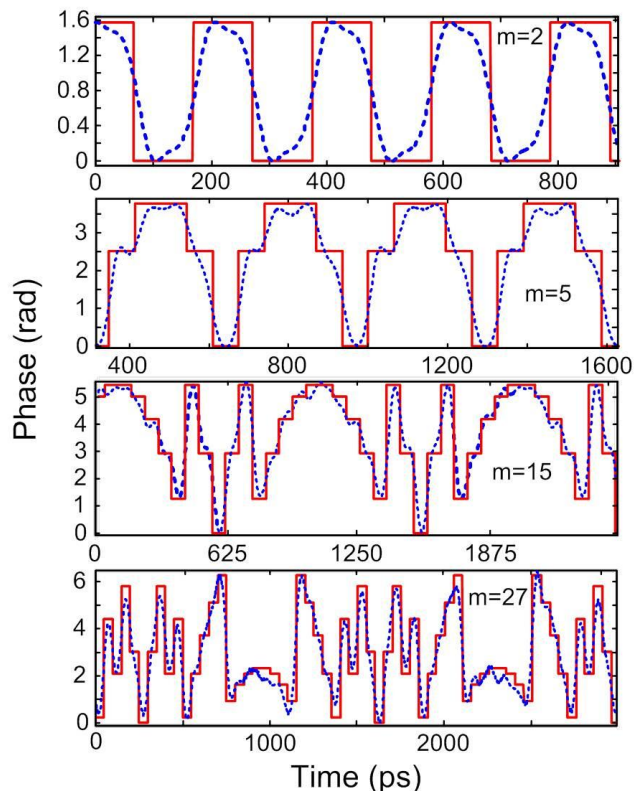

**Figure 2 | Experimental prescribed phase modulation profiles.** Temporal phase modulation patterns required for amplification factors $m$ =2, 5, 15, and 27, as determined by the Talbot carpet. The dashed blue curves show the experimental traces generated from an electronic arbitrary waveform generator (7.5-GHz analog bandwidth), as measured by an electrical sampling oscilloscope (40-GHz bandwidth), whereas the solid red curves show the ideal phase traces. For comparison, the experimentally measured electrical traces have been normalized to match the ideal phase profiles.

Figure 2 shows the prescribed electro-optic phase modulation profiles to the ~7-ps Gaussian input optical pulses generated by the fiber laser for the cases when we target gain factors of $m =$ 2, 5, 15, and 27, respectively. The temporal phase functions are generated from an electronic arbitrary waveform generator (AWG). The solid red lines show the ideal temporal phase profiles, and the dashed blue lines show the actual phase drive delivered by the AWG. Fig. 3 presents the optical spectra recorded with a high-resolution (20-MHz) optical spectrum analyzer after temporal phase modulation showing the predicted spectral Talbot effect, leading to the anticipated decrease in the frequency comb spacing by factors of 2, 5, 15, and 27, respectively. Fig. 4a shows the experimental results of the demonstrated passive amplification, with gain factors of $m =$ 2, 5, 15, and 27, at the output of the dispersive fiber link (total dispersion ~2,650ps/nm in the case of $m$=2 and $m$=5, and ~ 8,000ps/nm for $m$=15 and $m$=27). In the case of $m$=15 and $m$=27, fiber losses from dispersion are significant, ~42 dB from a total of six dispersion compensating fiber modules, so that we include active amplifiers, Erbium-doped fiber amplifiers (EDFAs), in our system in order to detect the signal at the end of the span to show proof-of-concept for these higher amplification factors. This practical limitation can be overcome by using lower loss dispersive devices, such as chirped fiber gratings, or using line-by-line spectral shaping with a linear waveshaper[29], for losses as low as 3 dB for the entire dispersive spectral phase employed[30].

The original repetition rates of the input pulse trains in the four reported experiments are 9.7 GHz, 15.43GHz, 15.43 GHz, and 19.75 GHz, respectively. The reduced rates after passive amplification are 4.85 GHz, 3.08 GHz, 1.03 GHz, and 0.73 GHz, as measured by the RF spectra of the detected output optical pulse trains, Fig. 5, in excellent agreement with the desired amplification factors. Fig. 4 shows the optical sampling oscilloscope (OSO) trace (500-GHz measurement bandwidth) of the pulse train after dispersion in the case with a phase-conditioned input (red) and without (blue). For $m = 2$, the pulse train intensity, as measured by the OSO, doubles as predicted. For $m = 5$, 15 and 27 the ideal amplification factors are nearly obtained, 98% of the desired $m =$5, 96% of the desired $m =$15, and 75% of the desired $m =$ 27. In all cases, the amplified temporal waveforms are nearly undistorted replicas of the unamplified Gaussian pulses. The reduction in the fidelity of the Gaussian spectral envelope (Fig. 3) for amplification factors $m =$ 15 and $m =$ 27, and the associated slight decrease of the expected gain, is mainly due to the time-resolution limitations of the AWG, which fails to reproduce the ideal temporal phase drive for more complicated phase patterns, Fig. 2. Fig. 4b further proves how Talbot passive amplification is achieved without affecting the temporal shape of the input waveform. Here, the laser pulses are re-shaped prior to amplification to exhibit a sinc-like pulse waveform in the time-domain. Using the same scheme that was applied to Gaussian pulses for amplification by 15 times, we similarly demonstrate passive amplification of the sinc-like pulses by a factor of ~12.6. The Talbot passive amplification approach can be applied on any arbitrary waveform with no fundamental limitation on the signal frequency bandwidth. It is only practically limited by the spectral bandwidth of the linear region of the particular dispersive medium. In the case of dispersive delay provided by optical fiber as employed in this work, linear dispersion around 1550 nm greatly exceeds the pulse bandwidth. For other wavelength regions or wave systems care may need to be taken to ensure a linear chirp provided by dispersive delay.

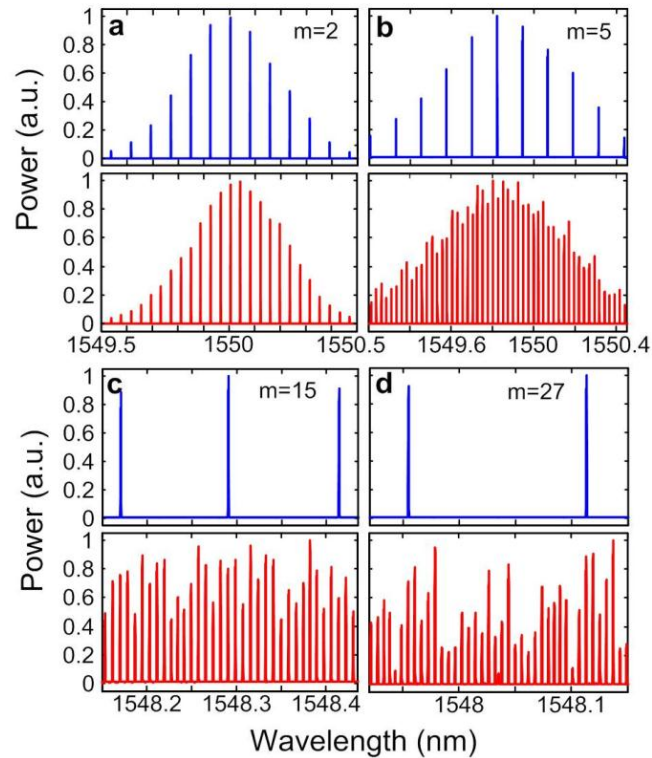

**Figure 3 | Measured optical spectra of the optical pulse trains.** Optical spectra of the optical pulse trains recorded with a high-resolution (20-MHz) optical spectrum analyzer after phase modulation with the phase modulator turned off (blue) and with the phase modulator turned on (red), demonstrating the expected decrease in the frequency comb spacing by factors of 2, 5, 15, and 27, respectively.

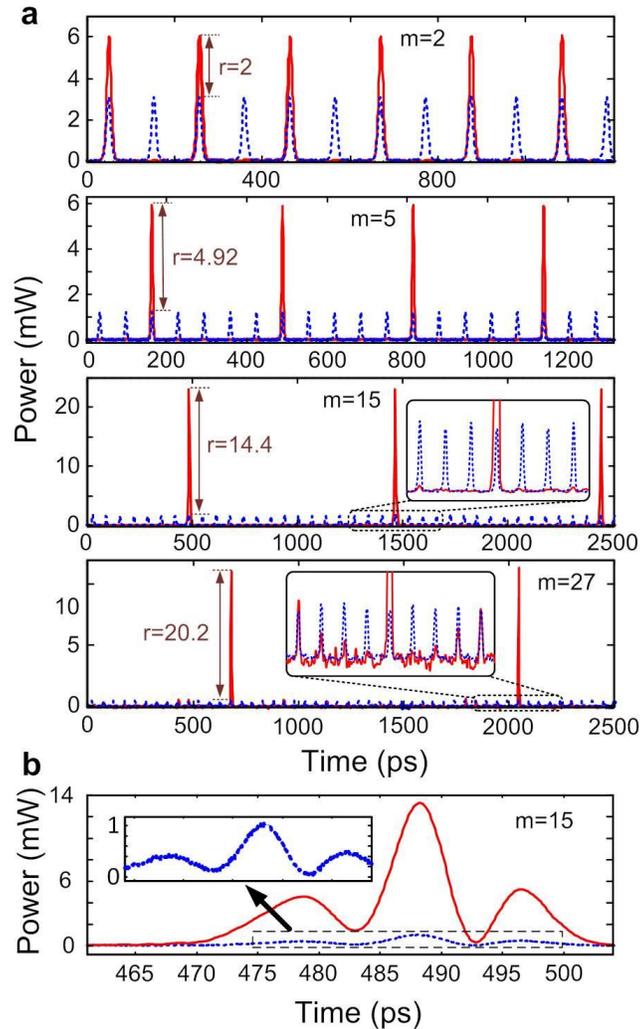

**Figure 4 | Experimental demonstration of passive waveform amplification**. (a) Optical sampling oscilloscope (500-GHz measurement bandwidth) time trace of pulse trains at the dispersive fiber output before passive amplification (dashed blue, with the phase modulator turned off) and after passive amplification (solid red, with the phase modulator turned on) for the desired amplification factors of $m = 2, 5, 15,$ and $27$. Experimental passive gain is measured as 2, 4.92, 14.4, and 20.2, respectively, on the optical sampling oscilloscope. (b) Passive amplification of an arbitrary waveform demonstrating the insensitivity of the Talbot method to temporal signal shape.

The plots in Fig. 6 demonstrate the phenomenon of passive amplification without the injection of intensity noise. In these experiments, no active amplification was used in the dispersive span. The identical match of the spectra, in the presence and absence of passive amplification, particularly at the noise-floor, indicates our technique does not contribute any measurable intensity noise, consistent with what one would expect from a passive system. Using the spectral linear interpolation method, the data show that the OSNR remains the same in the presence of amplification. To be more concrete, Fig. 6a shows the OSNR, in a 1 nm resolution bandwidth of the optical spectrum analyzer, to be 50 dB without passive amplification, PM-OFF (dashed blue), and 50 dB with passive amplification, PM-ON (red), indicating a noise figure of 0 dB. Within our ability to measure OSNR, we show there is no injected intensity noise from passive amplification. On the other hand, when an active amplifier (EDFA) is used at the end of the same network instead of passive amplification, the OSNR degrades from 50 dB to 36 dB, PM-OFF + EDFA (dot-dashed green), due to the injected ASE noise from the active amplifier. The injected noise from active amplification can also be inferred from the time traces shown in Figs. 6b. Although both active and passive amplification amplify the peak of the signal by approximately 7 dB (corresponding to passive gain of $m=5$ and the equivalent active gain of 5), the active amplification also raises the average noise floor by 7 dB (dot-dashed green) whereas passive amplification leaves the floor at its original level (red). Figure 6c shows optical oscilloscope traces with zero averaging, clearly confirming an increased fluctuation when active amplification is employed (top trace) as compared to passive amplification (bottom trace).

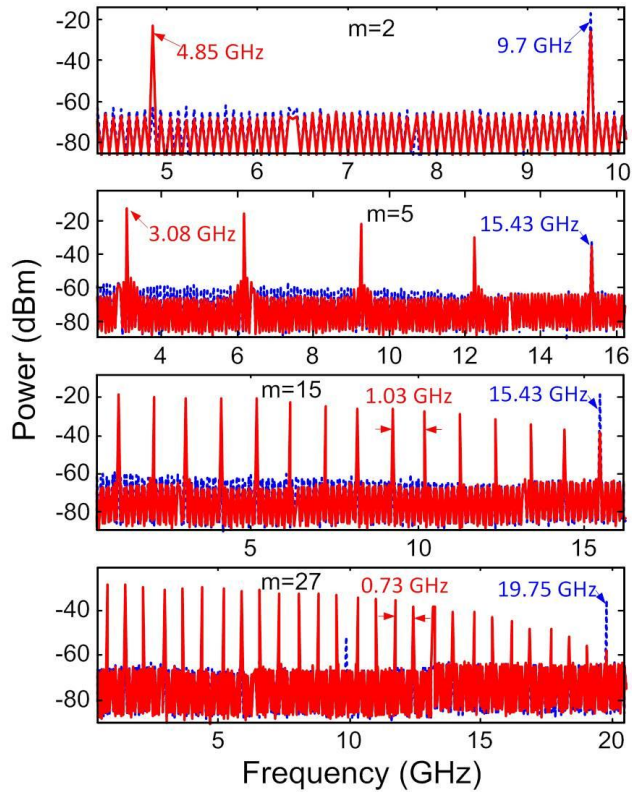

**Figure 5 | Experimental verification of repetition rate division for passive amplification.** Traces show the RF spectra of the optical pulse trains after photo-detection at the dispersive fiber output without passive amplification (dashed blue, with the phase modulator turned off) and with passive amplification (solid red, with the phase modulator turned on) for the desired amplification factors of 2, 5, 15 and 27.

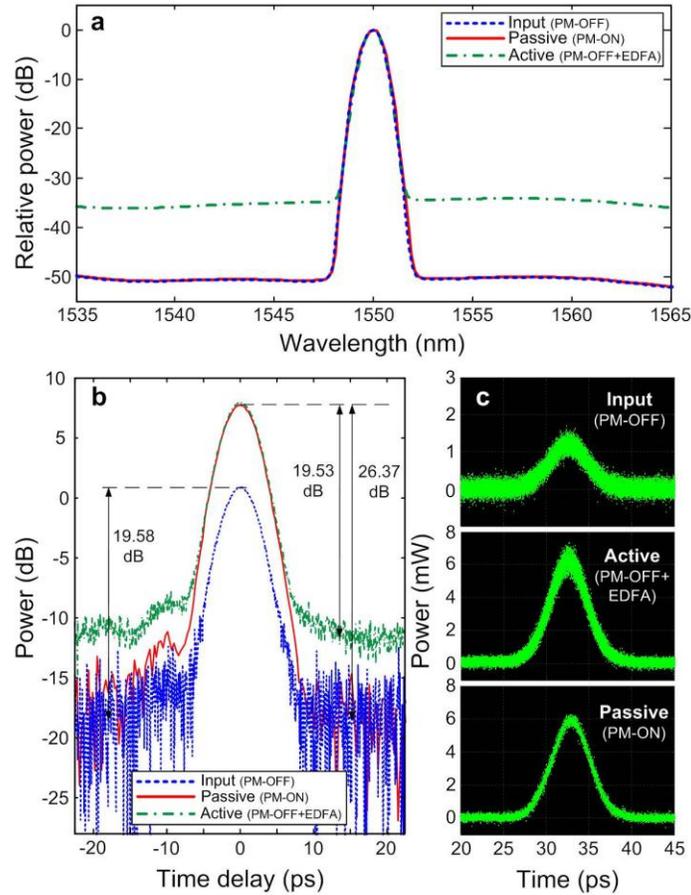

**Figure 6 | Experimental verification of noiseless amplification.** (a) Optical spectra of the pulse trains measured at the dispersive fiber output after passive amplification (phase modulator turned on, PM-ON) with *m*=5 (red), and before passive amplification (phase modulator turned off, PM-OFF), without (dashed blue) and with active amplification (EDFA) with gain of 5 (dot-dashed green). Measurement settings were fixed to a 1 nm resolution bandwidth for the optical spectrum analyzer. (b)-(c) Corresponding optical sampling scope time traces in averaging mode and sampling mode (no averaging), respectively.

Fig. 7 shows enhancement of extinction ratio (ER) as the passive amplification factor *m* increases for a noisy input with OSNR=10. In this experiment, ASE noise was injected onto the pulse train in a controlled fashion using an EDFA, paired with a variable attenuator, at the input of the phase modulator and placing two more EDFAs in the dispersive span. We define ER enhancement as the ratio of the ER of the Talbot-amplified signal with respect to the ER of the input train, where ER is given by the ratio of the average peak intensity of the waveform to the average intensity of the noise floor. Fig 7(a) shows simulated data (blue squares) of how the ER scales linearly with *m*, and experimental data points for *m*= 2, 5 and 15 overlaid (red triangles). The dashed blue line shows the expected linear trend for ER enhancement. Passive amplification increases the waveform peak intensity by $m$ times, while leaving the noise floor at its average value. After phase modulation, energy is distributed into $m$-times more frequency tones. After temporal redistribution by dispersive delay, the individual frequency components of the waveform will add coherently to give an average peak-power that is $m$-times the original signal. The noise floor, however, will remain the same. This is shown clearly in the optical sampling scope trace in Fig. 6b, as well as Fig. 7b and c, which show the optical sampling oscilloscope time traces with passive amplification (red) and without (dashed blue) for $m=5$ and $m=15$, respectively. Note that traces in 7b and 7c have been normalized to the same peak power rather than the noise floor to better show ER enhancement. So long as the new frequency components generated during phase modulation are the correct ones (dictated by a correct phase drive), the noise floor will have the same average level from the destructive interference as before. Intensity noise present at the top of the waveform signal will also keep the same average value through the passive amplification process, in such a way that the ER enhancement actually increases linearly by a factor lower than $m$ (approaching $m$ for a higher input OSNR).

Talbot amplification is not simply a time-domain equivalent of band-pass filtering (BPF) but does actually redistribute signal energy from in-band noise. In order to show this important feature of our technique, we additionally show data of Talbot amplification used in conjunction with band-pass filtering and compare it to Talbot amplification without band-pass filtering, and to band-pass filtering alone. Fig 7a shows the experimentally expected enhancement to ER when band-pass filtering is employed without Talbot-amplification (solid violet). Here the filter used had a nearly flat-top spectral response with a 40 dB bandwidth of 2.8nm. For this case of OSNR=10, Fig. 7a shows that BPF is better at enhancing the ER than passive amplification alone for $m=5$, and that it is equivalent for $m=15$. This can be also seen in the difference between the solid red and dot-dashed green curves in Fig. 7a and b. However, when a BPF is used in conjunction with Talbot amplification, the ER improves drastically since the BPF gets rid of the out-of-band noise, and the passive amplification additionally enhances the resulting ER as discussed above. In the case of $m=15$, the ER enhancement increases to 20.9 dB rather than ~11 dB from passive amplification alone. Likewise the ER enhancement for $m=5$ shifts to 17.7 dB rather than ~6.5 dB from passive amplification alone. We note that the ER enhancement from the BPF depends on the OSNR while the ER enhancement for Talbot amplification depends both on level of noise and amplification factor *m*. Talbot amplification becomes particularly effective for enhancing ER at low OSNRs whereas BPFs are less effective.

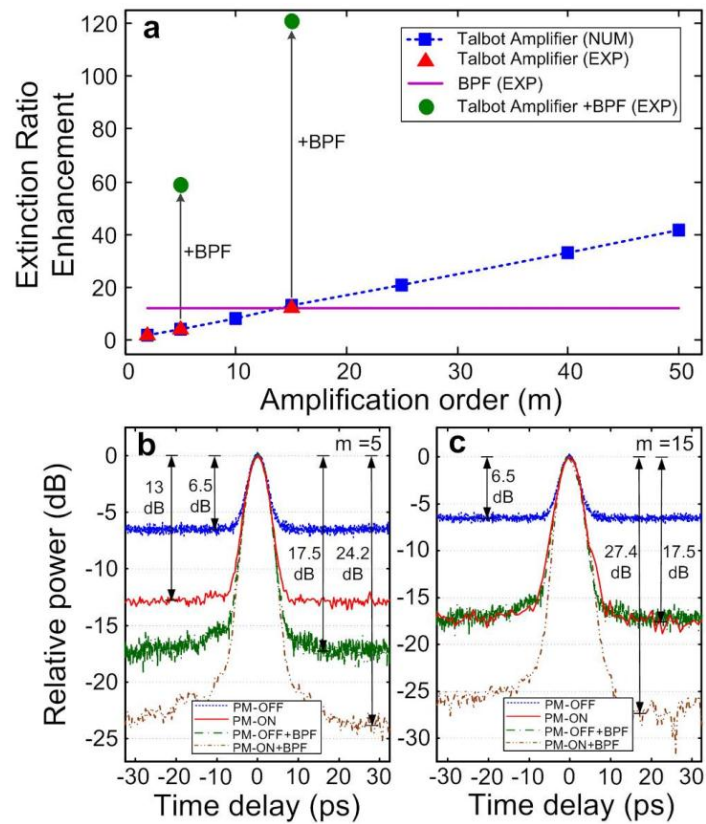

**Figure 7 | Extinction ratio enhancement.** (a) Enhancement of extinction ratio (ERE) as the passive amplification factor *m* increases for a noisy input with OSNR=10. The experimental data points (red triangles) for *m*=2, 5 and 15 are overlaid with the simulation trend (blue squares). The solid violet line shows the experimentally measured ERE when a band-pass filter is employed without Talbot-amplification. Solid green circles show experimental data points of ERE when a band-pass filter is used in conjunction with Talbot amplification. (b) Optical sampling oscilloscope time traces with passive amplification (red, PM-ON), without passive amplification (dashed blue, PM-OFF), with a BPF alone (dot-dashed green), and with both the use of a BPF and passive amplification (double-dot-dashed brown), for *m*= 5 (c) Similar optical sampling oscilloscope traces as in (b) but with *m*=15.

Fig. 8 shows how Talbot amplification behaves as a conventional averaging process, e.g., scope averaging, on ASE-like intensity noise fluctuations. Fig 8a shows experimental data for the coefficient of variance (CV), the ratio of the standard deviation to the mean for the top level, of a noisy pulse (OSNR=10) vs. the inverse of the square root of amplification factor *m*=N (red squares). Also shown is the CV vs. the inverse of the square root of number of scope averages N (blue circles), demonstrating the equivalence of Talbot amplification to averaging. The theoretical trendline for scope averaging, which scales as $\sqrt{N}$, is overlaid (dashed green). Experimental sampling oscilloscope traces in Fig. 8b and c show how the point-to-point fluctuation is nearly the same for scope averaging and Talbot amplification. Fig 8b (blue) shows results for a pulse without passive amplification and a regular scope average of N=15, and Fig. 8c shows results for a Talbot-amplified pulse by $m$=15 with no scope averaging. OSNR=10 for the experimental traces. Here the realignment of newly created frequencies with old ones, with just as much negative and positive fluctuation, creates an averaging effect. Because there are $m$-times as many frequency tones adding together, the reduction in fluctuation (square deviation of intensity noise) at the waveform top nearly follows the conventional counting rule and goes like the square root of $m$. Passive amplification is therefore equivalent to a *real-time optical average*. Said other way, Fig 8c is the equivalent of Fig 8b without the need for detection and post-processing. Such a real-time average could be particularly important where a clean pulse is needed directly in the optical domain. Notice also that whereas Talbot amplification is equivalent to scope averaging concerning its effect on white noise fluctuations, Talbot amplification additionally enhances the waveform train ER as the gain factor is increased. This is in sharp contrast to conventional scope averaging, where the noise floor and peak always average to their same respective levels, keeping ER constant.

Finally, Fig. 9 shows a side-by-side comparison of the oscilloscope time-trace of a Talbot amplified pulse train versus an actively amplified pulse train using an EDFA. Both amplification techniques amplified a noisy input pulse train with an input OSNR=5, and both had a gain of 15. The noisy input is shown in the top trace, and after active amplification, bottom trace, the pulse is significantly degraded. Noise is injected onto the already-noisy train, and both the noise floor and pulse peak are amplified about the same amount, negatively affecting the pulse quality. However, in the case of passive amplification, no noise is injected, the ER is enhanced by a factor approaching $m$- the peak is amplified while the noise floor is not, and the point-to-point fluctuations from ASE noise are reduced by a real-time optical average in which 15 averages have been taken. Both scope traces show zero averaging in the scope representation, but the Talbot amplified pulse has notably better extinction and less noise fluctuation. This signal recovery resembles the strategies used in optical spread-spectrum methods to hide an optical pulse in noise and recover it from the noisy background through the use of a pseudo-random spectral phase mask (transmitter) and its conjugate (receptor)[31]. However, for Talbot passive amplification the analogous "phase masks" are in time and frequency, respectively, instead of frequency alone, effectively implementing a coherent addition of multiple consecutive waveform copies.

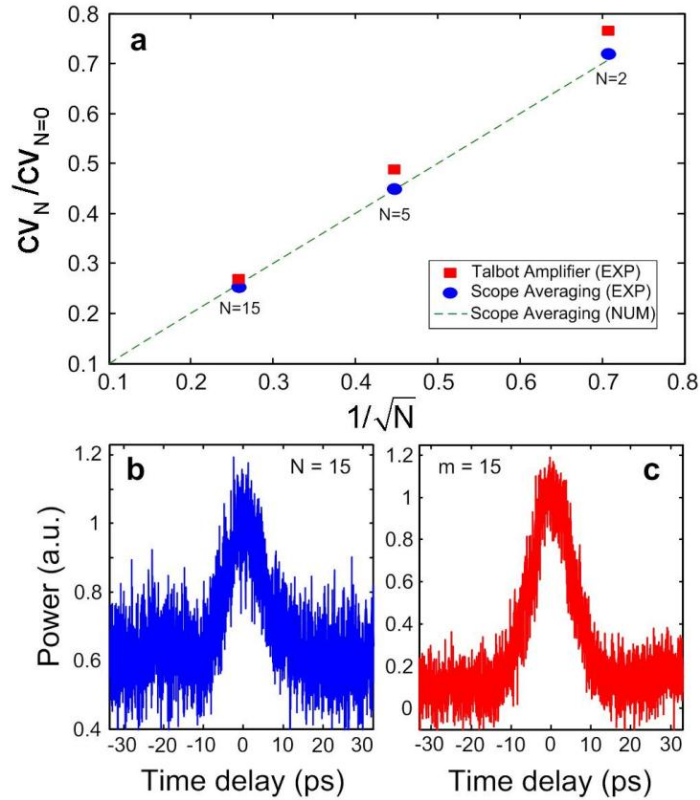

**Figure 8 | Averaging effect of Talbot amplification for ASE-like noisy fluctuations.** (a) Red squares show experimental data for the coefficient of variance (CV) of a passively amplified noisy pulse (OSNR=10) for a given passive amplification factor $m$ vs. the inverse of the square root of the amplification factor $m$ with no scope averaging; solid blue circles show experimental data for the CV of the same noisy input pulse (OSNR=10) for N scope averages and no passive amplification vs the inverse of the square root of the number of scope averages N; the dashed green line shows the expected $\sqrt{N}$ dependence of averaging governed by Poisson statistics. (b) Experimental sampling oscilloscope trace for a pulse without passive amplification using 15 scope averages, N=15 (c) Experimental sampling oscilloscope trace of the same pulse train as in (b) but using passive Talbot amplification with $m$=15 and no scope averaging.

Page **17** of **24**

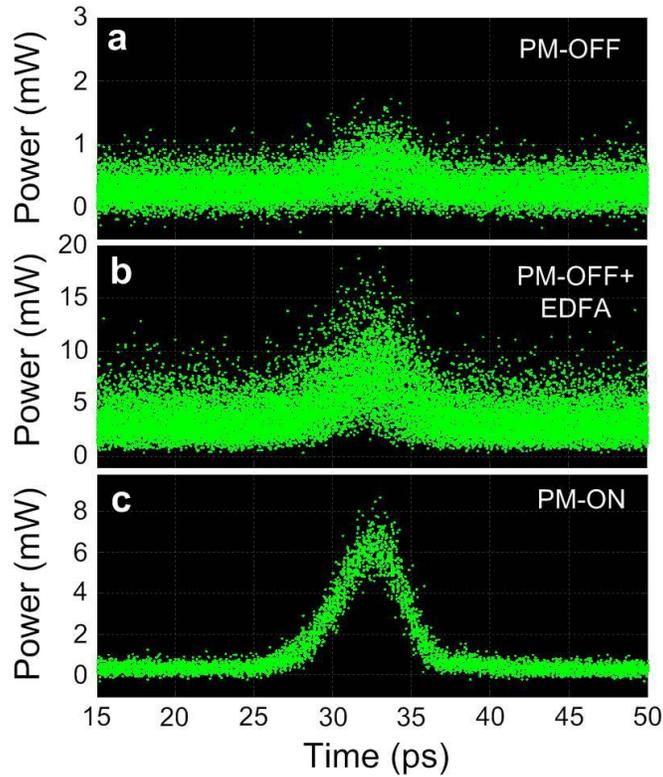

**Figure 9 | Comparison of passive and active amplification.** Oscilloscope time-traces of (a) a noisy pulse train (OSNR = 5) measured at the dispersive span output before passive amplification (PM-OFF), (b) Oscilloscope time-trace of the noisy signal shown in trace (a) after active amplification using an EDFA with a gain of 15, and (c) Oscilloscope time-trace of the noisy signal shown in trace (a) after passive Talbot amplification (PM-ON) with passive gain of 15. Oscilloscope is in sampling mode (no averaging) for all traces.

**Discussions and Conclusions**

Even though phase-only temporal and spectral manipulation processes ideally preserve the input signal energy, practical devices introduce losses, and these should be minimized in the design of amplification systems aimed at achieving input-to-output gain. Typically, highly-dispersive elements require large spectral phase shifts and these are typically achieved through long transmission paths, such as for the optical fiber solution shown here, which readily leads to loss. However, devices and materials can be engineered to simultaneously provide high levels of dispersion with high transmission. So long as the dispersive device provides linear dispersion over the bandwidth of the pulse train, there exist a variety of different options other than optical fiber with which to impose spectral phase. For example, in the optical region, chirped fibre Bragg gratings are available offering dispersion levels greatly exceeding 10,000ps/nm with less than 3dB loss[30]. Additionally, alternative methods exist for applying the spectral phase that is required for self-imaging *without* dispersive delay, such as spectral line-by-line shaping[29], for which spectral phase shifts can be restricted to the range $[0, 2\pi]$ *for any desired amplification factor.*

Though our technique cannot readily be used with aperiodic data, such as a random bit sequence, a single shot, or a sequence of data that is pulse-position coded, it can be made to work if such aperiodic data is made to repeat over some repetition period, even at a much slower rate. For example, the correct phase drive and subsequent dispersion could be engineered to amplify a repeating aperiodic sequence of 100 bits such that every set of 100 bits would coherently add. Alternatively, one could purposely modulate data at a faster rate in order to use Talbot-amplification to coherently add and amplify it at a slower desired output rate. Though our technique cannot generally be applied to data sequences, there are many important exceptions where it can be engineered to work.

Because dispersion in many wave systems can be controlled along wave propagation (as in the case of optical fibre links shown here), our method can also be used to tailor amplification to occur only at specific locations far away from any power source, offering the unique possibility of remote or localized waveform amplification. Perhaps of greater significance is how passive amplification using our method can be readily applied to other spectral regions or wave systems. Passive amplification by self-imaging is applicable to any wave system in which the temporal and dispersive spectral phase can be controlled. Such control is not only available throughout the entire electromagnetic spectrum[32], but also for many other wave systems, such as acoustic[33] and matter waves[25], for which active gain mechanisms are extremely challenging to implement or simply not available. As mentioned above, passive amplification of acoustic and vibrational waveforms could be useful in MEMS applications, sonar, and ultrasonic sensing and imaging. Additionally, passive probability amplification of matter-waves could provide another tool for atom optics to control and manipulate quantum states of matter[25] as well as single photon pulses[26].

In summary, we have proposed and experimentally demonstrated a new concept for waveform intensity amplification without using active gain by recycling energy already stored in the input signal. Our method uses dispersive self-imaging phenomena, temporal Talbot effects, which exploit the intrinsic repetitive nature of waveform signals to precisely redistribute the input signal energy into fewer waveforms in order to achieve noiseless amplification of each individual waveform. This technique allows us to overcome critical limitations of present active-gain based amplification methods. In particular, our demonstrated passive waveform amplification technique represents an amplification method that can be potentially applied to repetitive signals in all wave systems, eliminates the wasted power inherent in applications requiring peak-power amplification, does not amplify or inject noise in the output signal, and even enhances the extinction ratio and reduces noise fluctuation of the input waveform train.

**Methods**

We used a commercial actively mode-locked fibre laser (Pritel - Ultrafast Optical Clock) to generate a repetitive input optical pulse train with tunable repetition rate. Temporal pulses directly generated from the laser are Gaussian in shape, approximately 7.0 ps intensity FWHM. The appropriate multilevel temporal phase modulation required for a desired amplification factor was applied to the optical pulses with a commercial fibre-integrated electro-optic phase modulator (EOspace, 25 GHz bandwidth) driven by an electronic arbitrary waveform generator (Tektronix AWG7122C, 7.5 GHz analog bandwidth). Subsequent first-order dispersion was provided by six dispersion-compensating fibre modules (Corning PureForm DCM-D-080-04), with a total of ~8,000ps/nm for $m=15$ and $m=27$ and ~2,650ps/nm for $m=2$ and $m=5$. Each module had a loss of 7 dB. Therefore two active-gain erbium-doped fiber amplifiers (EDFAs) were used in the span for the cases of $m=15$ and $m=27$ to aid in signal detection after large fiber loss in order to show proof-of-concept of higher passive gain factors. Because the amount of dispersion required for coherent addition grows linearly with the desired amplification factor and quadratically with the temporal period of the repeating input waveform, changing the repetition rate of the input pulse train allowed us the experimental convenience of demonstrating different amplification factors at the output of the similar length of dispersive optical fibre link. Generation of optical pulses with a sinc-like temporal shape was achieved by filtering the input pulse train with a square-spectrum filtering function using an optical wave-shaper (Finisar 4000S).

For noise mitigation experiments, ASE noise was purposely injected onto the pulse train using an EDFA at the input of the phase modulator and placing two additional EDFAs in the dispersive span. Using a variable attenuator prior to the input EDFA, we controlled the input power, versus the fixed amount of injected ASE noise from the EDFAs in the system, thereby implementing a desired value of OSNR. For simulation, noise was modeled in the frequency-domain as random Gaussian fluctuation in both phase and amplitude, with mean spectral amplitude- and phase-noise components as zero and a standard deviation per frequency point normalized to sum to the desired average noise power, and corresponding desired OSNR.

Finally, all optical time-traces were recorded with a 500-GHz bandwidth optical sampling oscilloscope (Exfo PSO-100), the optical spectra were recorded with high-resolution (20-MHz) optical spectrum analyzer (Apex, AP2440A), the RF phase drive was recorded with a 40-GHz bandwidth electrical sampling oscilloscope (Tektronix CSA8200), and RF spectra were recorded by detecting the optical pulse train with a 45-GHz photo-detector (New Focus Model 1014) which was then input into a 26-GHz RF spectrum analyzer (HP 8563E).

**Acknowledgements:** We are gratefully indebted to Prof. Tudor W. Johnston for his careful reading of the manuscript. We thank Prof. Martin Rochette and Prof. Tayeb A. Denidni for lending the high-resolution optical spectrum analyzer and RF spectrum analyzer used in this work. We are also grateful to Mr. Robin Helsten for assistance with some of the reported experiments. This research was supported by the Natural Sciences and Engineering Research Council of Canada (NSERC), Le Fonds Québécois de la Recherche sur la Nature et les Technologies (FQRNT) and the Canada Research Chair in "Ultrafast Photonic Signal Processing". R.M acknowledges financial support from the Ministère de l'Éducation, du Loisir et du Sport du Québec through the Merit Scholarship Program for Foreign Students. J.V.H gratefully acknowledges INRS-EMT for support as a visiting researcher and Augustana College's Faculty Research Committee for travel assistance.

Correspondence and request for materials should be addressed to J.A. ([azana@emt.inrs.ca](azana@emt.inrs.ca)).